\documentclass[twocolumn,aps,prd,groupedaddress,showpacs,nofootinbib]
              {revtex4}
              \usepackage{mathrsfs}%
\usepackage{graphicx}
\usepackage{amsmath}
\usepackage{bm}
\usepackage{relsize}
\RequirePackage{xspace}
\newcommand{\be}{\begin{equation}}
\newcommand{\ee}{\end{equation}}
\newcommand{\bqa}{\begin{eqnarray}}
\newcommand{\eqa}{\end{eqnarray}}
\newcommand{\pslash}{\slash\hspace{-0.55em}}
\newcommand{\as}{\alpha_{\mathrm{s}}}

\begin{document}


\title{\mbox{}\\[10pt]
Puzzles in $B\rightarrow h_c(\chi_{c2})K$ Decays and QCD
Factorization}

\author{Ce Meng$~^{(a)}$, Ying-Jia Gao$~^{(a)}$, and Kuang-Ta Chao$~^{(b,a)}$}
\affiliation{ {\footnotesize (a)~Department of Physics, Peking
University,
 Beijing 100871, People's Republic of China}\\
{\footnotesize (b)~China Center of Advanced Science and Technology
(World Laboratory), Beijing 100080, People's Republic of China}}



\date{\today}

\begin{abstract}
We study the factorization-forbidden decays $B\to h_c K$ and
$B\to\chi_{c2}K$ in the QCD factorization approach. If neglecting
the vertex corrections and regularizing the end-point
singularities in spectator corrections properly, we get small
branching ratios for both the two decay modes, which are roughly
consistent with the experimental upper limits. This is in contrast
to another factorization-forbidden decay $B\to \chi_{c0}K$, for
which a large decay rate is obtained in the same approach.

\end{abstract}



\maketitle


$1.~Introduction.$ The branching ratios(BRs) of
$B\!\!\rightarrow\!\!h_c(\chi_{c0},\chi_{c2})K$ decays are expected
to be small since they are naively
factorization-forbidden~\cite{BSW}. But non-factorizable
contributions to these modes could be large and enhance their BRs.
This is the case for $B\! \rightarrow\!\chi_{c0}K$, which was
observed by Belle~\cite{Belle02} with surprisingly large BR of
$(6.0^{+2.1}_{-1.8}\pm 1.1)\times10^{-4}$ and was soon confirmed by
BaBar~\cite{BaBar04} with BR of $(2.7\pm0.7)\times10^{-4}$, being
comparable to those of factorization-allowed but color-suppressed
decays, such as $B\!\rightarrow\! J/\psi(\eta_c,\chi_{c1}...) K$.

One may then expect similarly large BRs for $B\rightarrow
h_c(\chi_{c2}) K$ decays. Accordingly, a number of authors have
suggested that $h_c$ could be easily observed in
B-factories~\cite{Gu}. However, recent measurements set the upper
limit BRs of $B\rightarrow \chi_{c2} K$~\cite{BaBar05} and
$B\rightarrow h_c K$~\cite{Belle06} at 90\% C.L.:
\begin{eqnarray}
{\mathrm{Br}} (B^{+}\rightarrow \chi_{c2} K^+)&<&
3.0\times10^{-5}, \nonumber\\
{\mathrm{Br}} (B^{0}\rightarrow \chi_{c2} K^{0}) &<&
4.1\times10^{-5}~, \nonumber\\
{\mathrm{Br}} (B^{+}\rightarrow \chi_{c2} K^+)&<& 3.8\times10^{-5},
\label{chic2hcex}
\end{eqnarray}
which are about an order of magnitude smaller than that of
$B\rightarrow \chi_{c0} K$. The puzzling hierarchies challenge our
understanding of the properties of the non-factorizable mechanism
for these decay modes.

The non-factorizable contributions to these modes have been studied
in different approaches. The final-state soft re-scattering effects
were suggested and large and comparable rates for all these three
modes were predicted~\cite{Cola}, which are evidently inconsistent
with new experimental data. On the other hand, with the Light-Cone
Sum Rules~\cite{Melic} the non-factorizable contributions were found
to be too small to account for the large $B\rightarrow \chi_{c0} K$
decay rate. Recently, $B\!\rightarrow\! \chi_{c0}K^{(*)}$ were also
studied in the PQCD approach~\cite{Li}, where the $k_T$
factorization was adopted. They predicted a large rate of
$B\rightarrow \chi_{c0} K$ by including the spectator contributions
only, but no predictions for rates of $B\rightarrow h_c(\chi_{c2})
K$ were given.

Within the framework of QCD factorization~\cite{BBNS},
$B\!\rightarrow\! \chi_{c0,2} K$ decays were studied
earlier~\cite{Chao03,Chao04}, and we found that there exist infrared
divergences in the vertex corrections and end-point singularities in
the leading twist spectator corrections, which implies that soft
contributions may be large in these decays. These contributions come
from the soft gluon exchange between the $K$ or $B$ meson and the
$c\bar c$ pair which is emitted as a color-octet at the
short-distance weak interaction vertex. This may imply a connection
of the infrared behavior in the exclusive decays with the
color-octet contributions in the inclusive B decays to charmnium ,
which are found to be the dominant mechanism~\cite{Beneke99}. The
results of Ref.~\cite{Beneke99} suggest that a sizable fraction of
the large color-octet partial rate of inclusive $B$ decay into
charmonium does in fact end up in some two-body decay modes.~This is
even true for the factorization-allowed decays, such as
$B\!\!\rightarrow\!\! J/\psi K$, for which the infrared safe
leading-twist contributions only result in a BR about $1\times
10^{-4}$, which is an order of magnitude smaller than the
experimental data~\cite{Cheng}.~However, after considering the soft
contributions arising from higher-twist spectator interactions, the
prediction can account for the data quite well~\cite{Cheng}.~The
prominent effects of soft spectator interactions are also used in
Ref.~\cite{Gao06} to explain the large BR of $B\!\rightarrow\!
\psi(3770) K$ decay.

Similarly, these soft contributions to $B\!\rightarrow\! \chi_{c0}
K$ were also estimated by the authors of Ref.~\cite{pham} and by
us~\cite{Meng0502} in QCD factorization. Both they and we found that
there were linear singularities in chirally enhanced twist-3
spectator interactions, which are numerically large and dominant in
$B\!\rightarrow\! \chi_{c0} K$ decay, but the results of theirs and
ours were quite different because different treatments of the
end-point singularities in the spectator interactions. They also got
a small BR for $B\!\rightarrow\! \chi_{c2} K$ by introducing
undetermined imaginary parts for soft spectator interactions.

Since the $s(\bar{s})$ quark emitted from the weak interaction
vertex moves fast in the B meson rest frame, we may expect that the
soft gluon exchange is dominated by that between the $c\bar c$ pair
and the spectator quark that goes into the kaon. For
$B\!\rightarrow\! \chi_{c0} K$ decay, this fact has been confirmed
in QCD factorization~\cite{pham,Meng0502} and PQCD
factorization~\cite{Li} approaches. In particular,
in~\cite{Meng0502} we find that the vertex corrections could be
small, and the spectator corrections are large and dominant. In this
letter, we will evaluate the BRs of $B\!\rightarrow\!h_c(\chi_{c2})
K$ by using the same schemes given in~\cite{Meng0502}. We will show
that if the contributions from vertex corrections are neglected, we
can fit the experimental data of $B\!\rightarrow\!h_c(\chi_{c2,0})
K$ quite well.

$2.~Amplitudes~in~QCD~factorization.$ We treat the charmonium as a
non-relativistic $c\bar c$ bound state. Let $p$ be the total
momentum of the charmonium and $2q$ be the relative momentum between
$c$ and $\bar c$ quarks, then $v^2 \sim 4q^2/p^2 \sim 0.25$ can be
treated as a small expansion parameter. For the P-wave charmonium,
because the wave function at the origin $\mathcal{R}_1(0)\!=\!0$,
the amplitudes need to be expanded to the first order in the
relative momentum $q$ (see, e.g.,~\cite{Kuhn}),
\begin{eqnarray}
 \label{amp}
\mathcal{M}(B\to\!\! {}^{2S+1}\!P_J(c\bar
c))\!=\!\!\!\!\sum_{L_z,S_z}\!\langle 1L_z;SS_z|JJ_z\rangle
 \!\int\!\!\frac{\mathrm{{d}}^4 q}{(2 \pi)^3}q_\alpha \nonumber\\
 \times \delta\!(q^0)\psi_{1M}^\ast\!(q)
 \mathrm{Tr}[\mathcal{O}^\alpha\!(0)P_{SS_z}\!(p,\!0)
\!+\!\mathcal{O}\!(0)P^\alpha_{SS_z}\!(p,\!0)],
 \end{eqnarray}
where $\mathcal{O}(q)$ represents the rest of the decay matrix
elements and are expected to be further factorized as product of $B
\to K$ form factors and hard kernel or as the convolution of a hard
kernel with light-cone wave functions of B meson and K meson within
QCD factorization approach. The spin projection operators
$P_{SS_z}(p,q)$ are constructed in terms of quark and anti-quark
spinors as
\begin{eqnarray}
 \label{spinp}
 P_{00}(p,q)&\!=\!\!&\sqrt{\!\frac{3}{m_c}}\!\sum_{s_1,s_2}\!\!v(\frac{p}{2}\!-\!q,s_2)
 \bar u(\!\frac{p}{2}\!+\!q,s_1) \!\langle s_1;\!s_2|00\rangle\nonumber\\
&\!\!=\!\!&-\sqrt{\frac{3}{4 M^3}}(\frac{\pslash p}{2}-\pslash
q-\frac{M}{2})\gamma_5(\pslash
p+M)\nonumber\\
 & &\times(\frac{\pslash p}{2}+\pslash q+\frac{M}{2}) ,\nonumber\\
P_{1S_z}(p,q)&\!=\!\!&\sqrt{\!\frac{3}{m_c}}\!\sum_{s_1,s_2}\!\!v(\frac{p}{2}\!-\!q,s_2)
 \bar u(\!\frac{p}{2}\!+\!q,s_1) \!\langle s_1;\!s_2|1S_z\!\rangle\nonumber\\
&\!\!=\!\!&-\sqrt{\frac{3}{4 M^3}}(\frac{\pslash p}{2}-\pslash
q-\frac{M}{2})\pslash \epsilon^\ast(S_z)(\pslash
p+M)\nonumber\\
 & &\times(\frac{\pslash p}{2}+\pslash q+\frac{M}{2}) ,
 \end{eqnarray}
and
 \bqa
\mathcal{O}^\alpha(0)\!&=&\!\frac{\partial \mathcal{O}(q)}{\partial
q_\alpha}|_{q=0},\\
P^\alpha_{SS_z}(p,0)\!&=&\!\frac{\partial P_{SS_z}(p,q)}{\partial
q_\alpha} |_{q=0}. \eqa
In Eq. (\ref{spinp}) we take charmonium mass $M\simeq 2 m_c$. Here
$m_c$ is the charm quark mass.

The integral in Eq.(\ref{amp}) is proportional to the derivative of
the P-wave wave function at the origin by
 \bqa
\int\!\frac{\mathrm{{d}}^3 q}{(2 \pi)^3}q^\alpha \psi_{1M}^\ast(q)
=i\varepsilon^{\ast\alpha}(L_z)\sqrt{\frac{3}{4\pi}}
\mathcal{R}^{'}_1(0), \eqa
where $\varepsilon^\alpha(L_z)$ is the polarization vector of an
angular momentum-1 system and the value of $\mathcal{R}^{'}_1(0)$
for charmonia can be found in, e.g., Ref.~\cite{Quig}.

We use the following polarization relations respectively for
$h_c(J=1)$ and $\chi_{c2}(J=2)$:
 \bqa\label{pol}
 \sum_{L_Z
S_Z}\!\!\varepsilon^{\ast\alpha}\!(\!L_z\!) \langle
1L_z;00|1J_z\rangle\!&=&\!\varepsilon^{\ast\alpha}(J_z),\nonumber\\
\sum_{L_Z
S_Z}\!\!\varepsilon^{\ast\alpha}\!(\!L_z\!)\epsilon^{\ast\beta}\!(\!S_z\!)
\langle
1L_z;\!1S_z|2J_z\rangle\!&=&\!\epsilon^{\ast\alpha\beta}(J_z).
 \eqa
where $\varepsilon^{\alpha}(J_z)$ is the polarization vector for
$h_c$, and the polarization tensor $\epsilon^{\alpha\beta}(J_z)$,
which is symmetric under the exchange
$\alpha\leftrightarrow\!\beta$, is the one appropriate for
$\chi_{c2}$ .

The effective Hamiltonian relevant for $B \to\! h_c(\chi_{c2})K$ is
written as~\cite{BBL}
 \bqa
\mathcal{H}_{\mathrm{eff}}\!\!=\!\!\frac{G_F}{\sqrt{2}} \Bigl(\!
V_{cb} V_{cs}^*\!(C_1 {\cal O}_1\!+C_2 {\cal O}_2 )\!-V_{tb}
V_{ts}^* \sum_{i=3}^{6} C_i {\cal O}_i \!\Bigr).
 \eqa
where $G_F$ is the Fermi constant, and $C_i$'s are the Wilson
coefficients. The relevant operators ${\cal O}_i$'s are given in
Ref.~\cite{Meng0502}.

\begin{figure}[t]
\vspace{-2.7cm} \hspace*{-2.9cm}
\includegraphics[width=14cm,height=16cm]{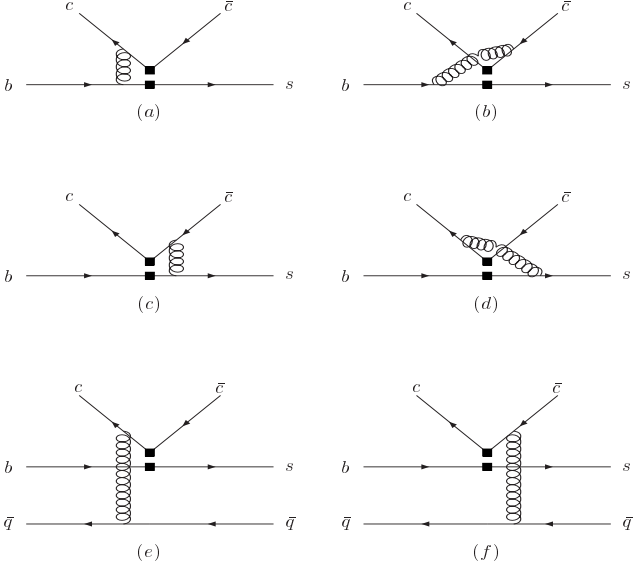}
\vspace{-4.8cm}
\caption{ Feynman diagrams for vertex and spectator corrections to
$B \to\! h_c(\chi_{c2}) K$.} \label{fvs}
\end{figure}

According to~\cite{BBNS} all non-factorizable corrections are due to
the diagrams in Fig.\ref{fvs}.  These corrections, with operators
${\cal O}_i$ inserted, contribute to the amplitude $\mathcal{O}(q)$
in (\ref{amp}), where the external lines of charm and anti-charm
quarks have been truncated. Taking non-factorizable corrections in
Fig.\ref{fvs} into account, the decay amplitudes for $B\to\!
h_c(\chi_{c0,2})K$ in QCD factorization can be written as
 \bqa\label{amp2}
  i\mathcal{M}_j =\frac{G_F}{\sqrt{2}}\Bigl[V_{cb}
V_{cs}^* C_1-V_{tb} V_{ts}^* (C_4 \mp C_6) \Bigr]\times A_j,
 \eqa
where in the parentheses, "$-$" is corresponding to $h_c$ and "+"
to $\chi_{c2}$, respectively. Here and afterwards, we will use the
subindex $j$ to denote different charmonium states, and $j\!=\!1,
2, $ represent $h_c$ and $\chi_{c2}$ respectively. Thus the
coefficients $A_j$'s are given by
\bqa\label{a} A_j=\frac{i 6 \mathcal{R}^{'}_1(0) }{\sqrt{\pi
M_j}}\cdot\frac{\alpha_s}{4\pi}\frac{C_F}{N_c}\cdot
F_1(M_j^2)\cdot n_j \bigl( fI_j +   fI\!I_j \bigr).
 \eqa
Here $F_1$ is one of the $B \!\!\to\!\! K$ form factors and the
relation ${F_0(M_j^2)}/{F_1 (M_j^2)}\simeq\!
1-{M_j^2}/{m_B^2}$~\cite{Cheng} has been used to simplify the
expressions of the amplitudes. The function $fI_j$ is calculated
from the four vertex diagrams (a, b, c, d) and $fI\!I_j$ is
calculated from the two spectator diagrams (e, f) in Fig.\ref{fvs}.
The function $fI\!I_j$ receives contributions from both twist-2 and
twist-3 light-cone distribution amplitudes of the K meson, and we
will simply symbolize them as $fI\!I_j^2$ and $fI\!I_j^3$,
respectively.

The factors $n_j$ in (\ref{a}) are given by
\bqa\label{nfactor}
n_1 &=& 2\varepsilon^\ast\cdot p_b,\nonumber\\
n_2 &=&
-\frac{4\sqrt{z_2}\epsilon^\ast_{\mu\nu}p_b^{\mu}p_b^{\nu}}{(1-z_2)m_B}
~,
 \eqa
where $z_j\!=\!M_j^2/m_B^2\!\approx\! 4m_c^2/m_b^2$.

For $B\rightarrow h_c(\chi_{c2})K$ decays, just like $B\rightarrow
\chi_{c0}K$, the soft gluons couple to the charm quark pair through
color dipole interactions, which are proportional to the relative
momentum $q$ and give the leading order contributions both in
$1/m_b$ and NR expansion (see Eq. (\ref{amp})). As a result, there
are generally infrared divergences in $fI_j$, while $fI\!I_j^{2,\,
3}$ suffer from logarithmic and linear end-point singularities,
respectively. The infrared divergences in $fI_j$ can be regularized
by a gluon mass $m_g$ or by the binding energies
$b_j=M_j-2m_c$~\cite{pham} of charmonia,
 \bqa\label{f1j}
 fI_j \!&=&\! -\frac{8
 z(1\!-\!z\!+\!\ln{z})}{(1-z)^2}\ln{(\frac{m_g^2}{m_b^2})}\!+\!\mbox{finite terms},\nonumber\\
 &&\hspace*{-1.5cm}=\! -\frac{4
 z((1\!\!-\!z)(1\!\!+\!3z)\!\!+\!4z\ln{z})}{(1-z)^2}\ln{\frac{b_j}{M_j}}\!+\!\mbox{finite terms},
 \eqa
which are consistent with Ref.~\cite{Chao04,Meng0502} and
Ref.~\cite{pham}, respectively. Here and afterwards we will omit the
subscript of $z_j$ unless it is necessary.

To derive the functions $fI\!I_j$, we use the light-cone projector
of $K$ meson up to twist-3,
\bqa
   M_{\alpha \beta}^K(p\,')&=&
    \frac{i f_K}{4}\Bigl\{ \pslash{p\,'}\,\gamma_5\,\phi_K(y)\nonumber\\
    &&\hspace*{-1.5cm}- \mu_K\gamma_5 \bigl( \phi_K^p(y) - i\sigma_{\mu\nu}\,p\,'^\mu
    \frac{\partial}{\partial k_{2\nu}}
    \frac{\phi_K^\sigma(y)}{6} \bigr) \Bigr\}_{\alpha\beta}\,,
\label{Kprojector}
  \eqa
where $k_{2(1)}$ is the momentum of the anti-quark (quark) in $K$
meson, and the derivative acts on the hard-scattering amplitudes in
the momentum space only. The chirally enhanced mass scale $\mu_K\!
=\! {m_K}^2/(m_s(\mu)+m_d(\mu))$ is comparable to $m_b$, which
ensures that the twist-3 spectator interactions are numerically
large, though they are suppressed by $1/m_b$.\footnote{In fact, the
twist-3 contributions to these decay modes are not power suppressed
because of the linear singularities contained in them.} The twist-2
light-cone distribution amplitude (LCDA) $\phi_K(y)$ and the twist-3
ones $\phi_K^p(y)$ and $\phi_K^\sigma(y)$ are symmetric under
exchange $y\!\leftrightarrow\!(1\!-\!y)$ in the limit of SU(3)
symmetry. In practice, we choose the asymptotic form $\phi_K
(y)\!=\!\phi_K^\sigma(y)\!=\!6y(1\!-\!y)$ and $\phi_K^p (y)\!=\!1$
for these LCDAs.

The projectors in Eq.~(\ref{Kprojector}) have been used by us in
Ref.~\cite{Meng0502} and there exist logarithmic and linear
end-point singularities in twist-2 and twist-3 spectator
interactions, respectively. Similar singularities are also found by
the authors of Ref.~\cite{pham}. But they use a different projector,
which can be derived from Eq.~(\ref{Kprojector}) by adopting an
integration by parts on $y$ and dropping the boundary terms. As
having mentioned in Ref.~\cite{Meng0502}, the two projectors can be
consistent with each other only when the singularities in all the
regions with small-virtualities are carefully regularized. That is,
to introduce a relative off-shellness $\lambda$ to regularize every
factor "$y$" in the denominator. For linear singularities, we have

\be\label{endpoint} \int \frac{y^n dy}{(y+\lambda)^{n+2}} =
\frac{1}{(n+1)\lambda}-1+O(\lambda),~n=0,1,2... \ee
The difference between Eq.~(\ref{endpoint}) in our scheme for
small-virtuality  and that in Ref.~\cite{pham} is quite evident. If
simply parameterizing the linear singularities on the left hand side
of Eq.~(\ref{endpoint}) as $\int dy/y^2=m_B/\Lambda_h$ following
Ref.~\cite{pham}, one would get the same results for all $n\geq 0$.
We will see that the non-trivial factors $(n+1)^{-1}$ in
Eq.~(\ref{endpoint}) play an important role in evaluating the BRs of
$B\!\rightarrow\!h_c(\chi_{c2})K$ decays.

Under this scheme, as we expected, the projectors in
Eq.~(\ref{Kprojector}) and in Ref.~\cite{pham} give the same forms
of $fI\!I_j^{2,3}$:
\bqa \label{f22final}
fI\!I_1^2\!\!\!&=&\!\!\!a_1\!\cdot\!\frac{m_B}{\Lambda_B}[3z+O(\lambda)],\nonumber\\
fI\!I_2^2\!\!\!&=&\!\!\!a_2\!\cdot\!\frac{m_B}{\Lambda_B}[6z\ln\lambda+15z\!+O(\lambda)],
 \eqa
\bqa \label{f23final}
fI\!I_1^3\!\!\!&=&\!\!\!\frac{a_1r_K}{1-z}\!\cdot\!\frac{m_B}{\Lambda_B}[\frac{-3z}{2\lambda}
\! - \!(1\!+\!z)\ln\lambda-\frac{3}{2}(1-z)+O(\lambda)],\nonumber\\
fI\!I_2^3\!\!\!&=&\!\!\!\frac{a_2r_K}{1-z}\!\cdot\!\frac{m_B}{\Lambda_B}[\frac{-\!z}{\lambda}
\! -
\!(1\!\!+\!2z)\ln\lambda\!\!-\!\!\frac{1}{2}(3\!+\!2z)\!+\!O(\lambda)],
 \eqa
where $r_K=2\mu_K/m_b$ is of $O(1)$ and  the factors $a_j$ are
defined as
\begin{equation}\label{afactor}
 a_j=\frac{8\pi^2 f_K f_B}{N_c(1-z_j)^2 m_B^2F_1(M_j^2)}.
 \end{equation}
In Eq.~(\ref{f22final}) and Eq.~(\ref{f23final}), the integral with
LCDA of B meson is conventionally parameterized as $\int_0^1 d\xi
\frac{\phi_B(\xi)}{\xi}=\frac{m_B}{\Lambda_B}$ with
$\Lambda_B\approx 300$ MeV~\cite{BBNS}.

It is worth emphasizing that this scheme is physical since the
off-shellness of quarks and gluons are naturally serve as infrared
cutoffs when $y\!\!\rightarrow\!0$. Following Ref.~\cite{Meng0502},
we use the binding energy $b_j$ to determine the relative
off-shellness $\lambda_j$:
\begin{equation}\label{lambdaj}
\lambda_j\simeq \frac{z_j}{1-z_j}\cdot\frac{b_j}{M_j}.
\end{equation}

Before explicitly evaluating these amplitudes, we should emphasize
the importance of the twist-3 spectator contributions again.
Normalizing all the amplitudes by the finite part of $fI_j$, we can
see that both $fI_j$'s and $fI\!I_j^2$'s are of the order
$\ln\lambda\sim\ln (m_B/\Lambda)$, but $fI\!I^3_j$'s are of order
$\frac{\mu_K}{\lambda m_B}\sim\frac{m_B}{\Lambda}$ because of the
chiral enhancement and the linear singularities contained in them.
These unusual power counting rules give support to the naive
expectation that the soft spectator interactions would be dominant
in these three decay modes. At the qualitative level, this statement
has been validated in Ref.~\cite{pham,Meng0502} for
$B\!\rightarrow\! \chi_{c0}K$.

$3.~Numerical~results~and~discussions.$ For numerical analysis, we
use the following input parameters with two values for the QCD
renormalization scale $\mu\!=\!1.45$ Gev and $\mu\!=\!4.4$ Gev:
 \bqa \label{parameter}
 &&\!\! M_1=3.524 ~\mbox{GeV}, ~ M_2=3.556 ~\mbox{GeV},~
\!m_c=1.5~\mbox{GeV}, \nonumber \\
&&\!\!  m_B\!=\!5.28~\mbox{GeV},~\!\! \Lambda_B\!=\!300~\!\mbox{MeV},~\!\!\mathcal{R}^{'}_1(0)\!=\!\sqrt{0.075}~\!\mbox{GeV}^{5/2}, \nonumber \\
&& \!\!f_B\!=\!216 ~\mbox{MeV}\mbox{\cite{Gray}}, ~
\!\!f_K\!=\!160~\mbox{MeV},\nonumber \\
&&\!\!F_1(M_1^2)=0.80,~F_1(M_2^2)=0.82~\mbox{\cite{Ball}},\nonumber \\
&&\!\!r_K(\mu) =0.85(1.3),~\as(\mu)=0.34(0.22).
 \eqa
In Eq.~(\ref{parameter}) the $\mu$-dependent quantities at
$\mu\!=\!1.45$ Gev ($\mu\!=\!4.4$ Gev) are shown without (with)
parentheses. The Wilson coefficients $C_i$'s are evaluated at
leading order by renormalization group approach~\cite{BBL}, since
the amplitudes in Eq.~(\ref{a}) are only of the leading order in
$\as$.

\begin{table}[t]
\caption{Theoretical predictions for functions $fI\!I_j$ and
${\mathrm{Br}}_j$. The values given at $\mu\!=\!1.45$ Gev
($\mu\!=\!4.4$ Gev) are shown without (with) parentheses.}
\label{f}\begin{tabular}
 {c|c|c|c|c}\hline
  &  \multicolumn{2}{c|}{$10^4\times {\mathrm{Br}}_j$} & $fI\!I_j^2$ & $ fI\!I_j^3(\mu)$
  \\\cline{2-3}
 $j$ & $m_g=0.5$ Gev & $m_g=0.2$ Gev &  & \\\hline
       $B\!\rightarrow\! h_c K$ & 1.6(0.8)  & 2.7(1.3) & 3.1  & $-$12.9($-$18.3) \\
  $B\!\rightarrow \!\chi_{c2}K$ & 0.8(0.3)  & 1.4(0.5) & 2.9  & $-$5.7($-$8.7) \\
  $B\!\rightarrow\! \chi_{c0}K$
  & 3.0(1.9)  & 4.2(2.4) & 7.8  & $-$35.0($-$53.6) \\ \hline
\end{tabular}
\end{table}

We use {\tt LoopTools}~\cite{looptools} for numerical calculations
of $fI_j$ with a gluon mass $m_g$, and the obtained BRs
${\mathrm{Br}}_j$ are listed in Tab.I with $m_g$ varying from 200
Mev to 500 Mev. The values of $fI\!I_j$ are also shown in Tab.I. We
see the $\mu$-dependence of $fI\!I_j^3$ arising from $r_K(\mu)$ is
largely cancelled by $\as(\mu)$~\cite{BBNS}. In Tab.I, for
comparison, we also list the results of $B\!\!\rightarrow\!\!
\chi_{c0}K$ decay, which can be found in
Ref.~\cite{Meng0502}\footnote{Here, the functions
$fI\!I_{\chi_{c0}}^{2,3}$ are extracted from Eq.~(\ref{a}) with
$n_{\chi_{c0}}\!=\!-(1-z)m_B/\sqrt{3z}$, so their definitions differ
from those in Ref.~\cite{Meng0502} just by a minus sign. }. Here and
afterward, we will use subscript "$h_c$, $\chi_{c2}$, $\chi_{c0}$"
to symbolize the quantities for $h_c$, $\chi_{c2}$ and $\chi_{c0}$
respectively.

We predict a relative small BR for $B\!\rightarrow\! \chi_{c2}K$
with
$R_{\chi_{c2}}\!\equiv\!{\mathrm{Br}}_{\chi_{c0}}/{\mathrm{Br}}_{\chi_{c2}}\!=\!3\!-\!6$
in Tab.I, which gives a signal of hierarchy without fine tuning of
parameters. However, the BR of $B\!\!\rightarrow\!\! h_cK$ is large
and the ratio
$R_{1}\!\equiv\!{\mathrm{Br}}_{\chi_{c0}}/{\mathrm{Br}}_{h_c}\!\simeq2$,
which is not large enough to account for experimental data.

Further, we regularize the infrared divergences in $fI_j (j=h_c,
\chi_{c2}, \chi_{c0})$ using the binding energy. The results for
$fI_{\chi_{c2}}$ and $fI_{\chi_{c0}}$ are available in
Ref.~\cite{pham}. We find both $fI_{\chi_{c2}}$ and $fI_{\chi_{c0}}$
in the binding energy scheme are smaller than those in the gluon
mass scheme (with the gluon mass values chosen above). Since the
large ${\mathrm{Br}}_{\chi_{c0}}$ are mainly due to the large
twist-3 spectator contribution
$fI\!I_{\chi_{c0}}^3$~\cite{Meng0502}, it is not very sensitive to
the value of $fI$, but ${\mathrm{Br}}_{\chi_{c2}}$ is. As a result,
we get $R_{\chi_{c2}}>20$ no matter which $\mu$ is chosen. Numerical
calculations indicate that the value of $fI_{h_c}$ in the binding
energy scheme is also smaller than that in the gluon mass scheme.

The largeness of $R_{\chi_{c2}}$ is mainly due to the difference of
$fI\!I_{\chi_{c2}}^3$ and $fI\!I_{\chi_{c0}}^3$, and can be traced
to the infrared behavior of the spectator interactions. This is the
reason why our results are different from those of Ref.~\cite{pham},
in which the coefficient of the linear pole in $fI\!I_{\chi_{c2}}^3$
is two times larger than ours in (\ref{f23final}). As a result, the
ratio $R_{\chi_{c2}}$ can be large only by adding adjusted imaginary
parts to spectator functions $fI\!I_j$ there.

\begin{table}[t]
\caption{Predictions of ${\mathrm{Br}}_j$ with only spectator
contributions $fI\!I_j$.}
\label{brf3only}\begin{tabular}{c|c|c} \hline
  $10^4\times {\mathrm{Br}}_j$ & $\mu=4.4$ Gev & $\mu=1.45$ Gev \\ \hline
  $B\rightarrow h_c K$      &  0.27 &  0.27 \\
  $B\rightarrow \chi_{c2}K$ &  0.03 &  0.02 \\
$B\rightarrow \chi_{c0}K$   &  1.05 &  1.17 \\\hline
\end{tabular}
\end{table}

As we have mentioned, since the $s(\bar{s})$ quark going out from
the weak interaction vertex moves fast in the B meson rest frame,
the soft gluon emitted from the $c\bar{c}$ quark pair is favored to
be absorbed by the spectator quark. This picture has been confirmed
by the power counting rules deduced above and by our analysis for
$B\!\!\rightarrow\!\! \chi_{c0}K$ decay~\cite{Meng0502}. On the
other hand, the vertex functions $fI_j$'s in general are numerically
small but sensitive to regularization schemes. So it might be
reasonable to evaluate amplitudes with only spectator contributions
$fI\!I_j^2$ and $fI\!I_j^3$, and treat vertex corrections as
uncertainties, of which the contributions to ${\mathrm{Br}}_j$ are
expected to be smaller than $1\times 10^{-4}$, as in the case of
factorizable decay $B\rightarrow J/\psi~K$. And it could be even
smaller for nonfactorizable decays. With the approximation of
neglecting contributions from vertex corrections, the results are
listed in Tab.II.

In this approximation, the soft spectator interactions dominate, and
there is no surprise that we get a larger
${\mathrm{Br}}_{\chi_{c0}}$ and a much smaller
${\mathrm{Br}}_{\chi_{c2}}$ simultaneously. The hierarchy is more
evident and $R_{\chi_{c2}}\!>\!30$. Furthermore, the rate of $h_c$
is  smaller than that in Tab.I, and $R_{h_c}\!\sim\!4$, which is
roughly consistent with the experimental data if we take into
account the new PDG average value
${\mathrm{Br}}_{\chi_{c0}}=(1.6^{+0.5}_{-0.4})\times 10^{-4}$.

Since soft scattering can introduce a strong-interaction phase, the
functions $fI\!I_j$ could be complex. In our scheme, the phases may
emerge with logarithmic singularities in $fI\!I_j$ if $\lambda$ is
negative. To account for this effect, here every logarithm  in
(\ref{f22final}) and (\ref{f23final}) can be multiplied by a
universal factor $(1\!+\!\rho e^{i\theta})$~\cite{BBNS} with
$0\!<\!\rho\!<\!1.5$. The phase $\theta$ is expected to be small,
say, varying from 0 to $\pi/2$. We find that the corrections to
those $BR_j$'s do not change our conclusions drastically. For
example, choosing $\rho\!=\!1.3$  with $\theta$ varying from 0 to
$\pi/2$, we get
${\mathrm{Br}}_{\chi_{c0}}\!\simeq\!(0.4-4.0)\!\times\!10^{-5}$,
$R_{\chi_{c2}}\!\simeq\!100$ and $R_{h_c}\!\simeq\!4$, which are
roughly consistent with experimental data, respectively.

In summary, since the non-factorizable contributions arising from
soft spectator interactions could be large for B to charmonium
exclusive decays, the smallness of experimental upper limits of BRs
of $B\!\to\! h_c(\chi_{c2})K$ are surprising. We study these two
modes within the framework of QCD factorization. If neglecting the
vertex corrections and regularizing the end-point singularities in
spectator corrections properly, we get small BRs of $0.27\times
10^{-4}$ and $0.03\times 10^{-4}$ for $B\to h_cK$ and
$B\to\chi_{c2}K$ respectively, while the predicted BR for
$B\!\to\!\chi_{c0}K$ is large, as shown in Table II. These are
roughly consistent with experiments. On the other hand, at present
we do not have a rigorous treatment for the vertex corrections
because of the appearance of infrared divergences. Nevertheless, if
we use the gluon mass (or binding energy) to regularize the infrared
divergences (with reasonable values for the gluon mass or binding
energy), we find their effects are possibly small, and may therefore
be viewed as theoretical uncertainties. If we take these
uncertainties into account, the predictions for BRs of $B\!\to\!
h_c(\chi_{c2})K$ become
${\mathrm{Br}}_{h_c}\!=\!(3\!\pm\!10)\!\times\!\!10^{-5}$ and
${\mathrm{Br}}_{\chi_{c2}}\!=\!(0\!\pm\!10)\!\times\!\!10^{-5}$.
Here the large errors in predictions  are mainly attributed to
uncertain but sub-important (even power-suppressed) contributions
arising from those vertex corrections.

$Note.~$ Very recently, the decay mode $B\!\to\! h_c K$ has also
been studied in PQCD factorization approach \cite{xqli}. As in the
case of $B\!\to\!\chi_{c0}K$~\cite{Li}, the authors of \cite{xqli}
neglected the vertex corrections, and get a small rate
${\mathrm{Br}}(B\!\to\! h_c K)=4.5\!\times\!10^{-5}$, which is
roughly consistent with ours.

\begin{acknowledgments}

We thank H.Y. Cheng for useful comments. This work was supported
in part by the National Natural Science Foundation of China (No
10421003) and the Key Grant Project of Chinese Ministry of
Education (No 305001).
\end{acknowledgments}



\begin{thebibliography}{}
\bibitem{BSW}
M.~Bauer, B.~Stech and M. Wirbel, Z. Phy. C {\bf 34}, 103 (1987)

\bibitem{Belle02}
K. Abe {\it et al.} (Belle Collaboration), Phys.\ Rev.\ Lett.\ {\bf
88}, 031802 (2002).



\bibitem{BaBar04}
B. Aubert {\it et al.} (BaBar Collaboration), Phys.\ Rev.\ D {\bf
69}, 071103 (2004).

\bibitem{Gu}
Y.F.~Gu, Phys. Lett. B {\bf 538}, 6 (2002); M. Suzuki, Phys.\
Rev.\ D {\bf 66}, 037503 (2002); T. Barnes, T.E. Browder, and S.F.
Tuan, hep-ph/0408081.


\bibitem{BaBar05}
B. Aubert {\it et al.} (BaBar Collaboration), Phys.\ Rev.\ Lett.
{\bf 94}, 171801 (2005); N. Soni {\it et al.} (Belle Collaboration),
Phys. Lett.  B {\bf 634 }, 155 (2006).

\bibitem{Belle06}
F. Fang {\it et al.} (Belle Collaboration), Phys.\ Rev.\ D {\bf 74},
012007 (2006) [hep-ex/0605007 v2].


\bibitem{Cola}
 P. Colangelo, F. de Fazio, and T.N. Pham, Phys.
Lett. B {\bf 542}, 71 (2002); Phys.\ Rev.\ D {\bf 69},  054023
(2004).
\bibitem{Melic}
 B. Meli\'c, Phys. Lett. B {\bf 591} 91, (2004);
Z.G.~Wang, L.~Li, T.~Huang, Phys.\ Rev.\ D {\bf 70},  074006 (2004).

\bibitem{Li}
 C.H. Chen  and H.N. Li,
Phys.\ Rev.\ D {\bf 71}, 114008 (2005).

\bibitem{BBNS}
M. Beneke, G. Buchalla, M. Neubert and C.T. Sachrajda, Phys.\ Rev.\
Lett. {\bf 83 }, 1914 (1999); Nucl. Phys. B {\bf 591 }, 313 (2000);
Nucl. Phys. B  {\bf 606 }, 245 (2001).



\bibitem{Chao03}
 Z.Z. Song and K.T. Chao, Phys. Lett.  B
{\bf 568 }, 127 (2003).

\bibitem{Chao04}
Z.Z. Song, C.~Meng, Y.J. Gao and K.T. Chao,  Phys.\ Rev.\ D {\bf
69}, 054009 (2004).

\bibitem{Beneke99}
M. Beneke, F. Moltoni and I.Z. Rothstein, Phys.\ Rev.\ D {\bf 59},
054003 (1999).

\bibitem{Cheng}
H.Y. Cheng and K.C. Yang, Phys.\ Rev.\ D {\bf 63}, 074011 (2001); J.
Chay and C. Kim, hep-ph/0009244.

\bibitem{Gao06}
Y.J. Gao, C. Meng and K.T. Chao,  Eur. Phys. J. A {\bf 28}, 361
(2006) [hep-ph/0606044].

\bibitem{pham}
 T.N. Pham and G.H. Zhu,
Phys. Lett.  B {\bf 619 }, 313 (2005).


\bibitem{Meng0502}
C.~Meng, Y.J. Gao and K.T. Chao,  hep-ph/0502240.


\bibitem{Kuhn}
J.H. K\"{u}hn, Nucl. Phys. B  {\bf 157}, 125 (1979); B. Guberina et
al., Nucl. Phys. B  {\bf 174}, 317 (1980).












\bibitem{Quig}
 E.J. Eichten and C. Quigg, Phys.\ Rev.\ D {\bf 52}, 1726 (1995).




\bibitem{BBL}
 G. Buchalla, A.J. Buras and M.E. Lautenbacher,
Rev. Mod. Phys. {\bf 68}, 1125 (1996).




\bibitem{Gray}
A. Gray {\it et al.}, Phys.\ Rev.\ Lett. {\bf 95}, 212001 (2005).



\bibitem{Ball}
P. Ball and R. Zwicky, Phys.\ Rev.\ D {\bf 71}, 014015 (2005);
hep-ph/0406261.

\bibitem{looptools}
T. Hahn and M. Perez-Victoria, Comput.\ Phys.\ Commun.\ 153, 118
(1999).

\bibitem{PDG2006}
W.M. Yao {\it et al.}. (Particle Data Group), J. Phys. G 33, 1
(2006)

\bibitem{xqli}
X.Q. Li, X. Liu and Y.M. Wang, hep-ph/0607009.



\end{thebibliography}
\end{document}